\documentclass[12pt]{article}

\usepackage{nature}

\newcommand{\msun}{{\mathrm M}_{\odot}}
\newcommand{\lsun}{{\mathrm L}_{\odot}}
\newcommand{\rsun}{{\mathrm R}_{\odot}}

\newcommand{\kms}{\mathrm{km}\,\mathrm{s}^{-1}}
\newcommand{\second}{\mathrm{s}}
\renewcommand{\day}{\mathrm{d}}
\newcommand{\yr}{\mathrm{yr}}
\newcommand{\myr}{\mathrm{Myr}}
\newcommand{\gauss}{\mathrm{G}}
\newcommand{\oned}{1D\@\xspace}

\newcommand{\threed}{3D\@\xspace}
\newcommand{\alfven}{Alfv\'en\@\xspace}
\newcommand{\tausco}{$\tau$~Sco\@\xspace}

\newcommand{\mesa}{\mbox{\textsc{Mesa}}\xspace}
\newcommand{\arepo}{\mbox{\textsc{Arepo}}\xspace}

\newcommand*{\eg}{e.g.\@\xspace}
\newcommand*{\ie}{i.e.\@\xspace}

\newcommand{\mytitle}{How mergers magnetise massive stars}
\title{\mytitle}

\newcommand{\mailf}{fabian.schneider@uni-heidelberg.de}
\newcommand{\mails}{sebastian.ohlmann@mpcdf.mpg.de}

\author[1,2,3$\ast$]{Fabian R.N. Schneider}
\author[4,2$\ast$]{Sebastian T. Ohlmann}
\author[3]{Philipp Podsiadlowski}
\author[2,5]{Friedrich K. R\"{o}pke}
\author[3]{Steven A. Balbus}
\author[6]{R{\"u}diger Pakmor}
\author[6]{Volker Springel}

\affil[1]{\normalsize{Zentrum f\"{u}r Astronomie der Universit\"{a}t Heidelberg, Astronomisches Rechen-Institut, M\"{o}nchhofstr. 12-14, 69120 Heidelberg, Germany}}
\affil[2]{\normalsize{Heidelberger Institut f\"{u}r Theoretische Studien, Schloss-Wolfsbrunnenweg 35, 69118 Heidelberg, Germany}}
\affil[3]{\normalsize{Department of Physics, University of Oxford, Keble Rd, Oxford OX1 3RH, United Kingdom}}
\affil[4]{\normalsize{Max Planck Computing and Data Facility, Gie\ss{}enbachstr.\ 2, 85748 Garching, Germany}}
\affil[5]{\normalsize{Zentrum f\"{u}r Astronomie der Universit\"{a}t Heidelberg, Institut f\"{u}r Theoretische Astrophysik, Philosophenweg 12, 69120 Heidelberg, Germany}}
\affil[6]{\normalsize{Max-Planck-Institut f{\"u}r Astrophysik, Karl-Schwarzschild-Str.\ 1, 85748 Garching, Germany}\vspace{0.5cm}}

\affil[$\ast$]{\normalsize{These authors contributed equally to this work.}}
\affil[ ]{\normalsize{To whom correspondence should be addressed; E-mail: \mailf\ and \mails.}}

\date{}

\makeatletter
\let\thetitle\@title
\let\theauthor\@author
\let\thedate\@date
\makeatother

\begin{document}


\baselineskip24pt


\maketitle

\begin{abstract}
Magnetic fields are ubiquitous in the Universe. The Sun's magnetic field drives the solar wind and causes solar flares and other energetic surface phenomena that profoundly affect space weather here on Earth. The first magnetic field in a star other than the Sun was detected in 1947 in the peculiar A-type star 78~Vir \cite{1947PASP...59..112B}. It is now known that the magnetic fields of the Sun and other low-mass stars (${\lesssim}\,1.5$ solar masses, $\msun$) are generated in-situ by a dynamo process in their turbulent, convective envelopes \cite{1979cmft.book.....P}. Unlike such stars, intermediate-mass and high-mass stars (${\gtrsim}\,1.5\,\msun$; referred to as ``massive'' stars here) have relatively quiet, radiative envelopes where a solar-like dynamo cannot operate. However, about 10\% of them, including 78~Vir, have strong, large-scale surface magnetic fields whose origin has remained a major mystery till today \cite{2009ARA&A..47..333D,2015A&A...582A..45F,2017MNRAS.465.2432G}. The massive star \tausco is a prominent member of this group and appears to be surprisingly young compared to other presumably coeval members of the Upper Scorpius association. Here, we present the first \threed magneto-hydrodynamical simulations of the coalescence of two massive main-sequence stars and \oned stellar evolution computations of the subsequent evolution of the merger product that can explain \tausco's magnetic field, apparent youth and other observed characteristics. We argue that field amplification in stellar mergers is a general mechanism to form strongly-magnetised massive stars. These stars are promising progenitors of those neutron stars that host the strongest magnetic fields in the Universe \cite{2009ARA&A..47..333D}, so-called magnetars, and that may give rise to some of the enigmatic fast radio bursts \cite{2017ApJ...841...14M}. Strong magnetic fields affect the explosions of core-collapse supernovae \cite{2014MNRAS.445.3169O} and, moreover, those magnetic stars that have rapidly-rotating cores at the end of their lives might provide the right conditions to power long-duration gamma-ray bursts \cite{2009MNRAS.396.2038B} and super-luminous supernovae \cite{2010ApJ...717..245K}.
\end{abstract}

It has been suggested that the strong magnetic fields observed in ${\approx}\,10\%$ of massive stars are inherited from a magnetised molecular cloud from which stars formed \cite{2012ARA&A..50..107L}. However, this mechanism cannot explain why only a rather small fraction of massive stars are magnetic. A selective process must be at work that gives rise to magnetic fields only in some stars and the merging of main-sequence (MS) stars and pre-MS stars has been hypothesised as an explanation \cite{2009MNRAS.400L..71F,2014MNRAS.437..675W}. The predicted fraction of massive merger products in the Milky Way is of order 10\% \cite{1992ApJ...391..246P,2014ApJ...782....7D} and could therefore explain the observed incidence of magnetic massive stars. The merger hypothesis is further supported by an apparent dearth of magnetic stars in close binaries \cite{2002A&A...394..151C,2015IAUS..307..330A}, which is to be expected if the mergers of close binaries produce some magnetic stars in the first place. Furthermore, merging can rejuvenate stars such that they appear younger than they really are \cite{1983Ap&SS..96...37H}. Rejuvenated stars are also known as blue stragglers. The aforementioned magnetic star \tausco likely belongs to this class, because, with an inferred age of ${<}\,5\,\myr$, it appears anomalously young compared to other ${\approx}\,11\,\myr$ old members of the Upper Scorpius association that are all thought to have formed together at about the same time. A merger origin of \tausco would naturally explain this discrepancy \cite{2016MNRAS.457.2355S}.

In the following, we show that \tausco's magnetic field and appearance as a blue straggler can indeed be understood as the result of the merger of two massive stars. To this end, we conduct---for the first time---\threed ideal magneto-hydrodynamical (MHD) simulations of the merger of a $9\,\myr$ old binary consisting of a $9\,\msun$ and a $8\,\msun$ core-hydrogen burning star with the moving-mesh code \arepo \cite{2010MNRAS.401..791S}, which is ideally suited for such simulations (see Methods). The binary configuration and evolutionary stage are chosen such that the resulting merger product is expected to have a total mass similar to \tausco (${\approx}\,17\,\msun$) and that the binary could have formed at the same time with other stars in the Upper Scorpius association. After the coalescence, we continue the evolution of the merger product in the \oned stellar evolution code \mesa \cite{2011ApJS..192....3P} and can thus follow the detailed dynamics of the coalescence and the subsequent thermal and nuclear evolution of the merger product up to the current state of \tausco and beyond (see Methods and Extended Data Figs.~\ref{fig:tvisc-tcool} and~\ref{fig:3d-vs-1d}).

\begin{figure}
\centering
\vspace{-2cm}
\includegraphics[width=0.92\textwidth]{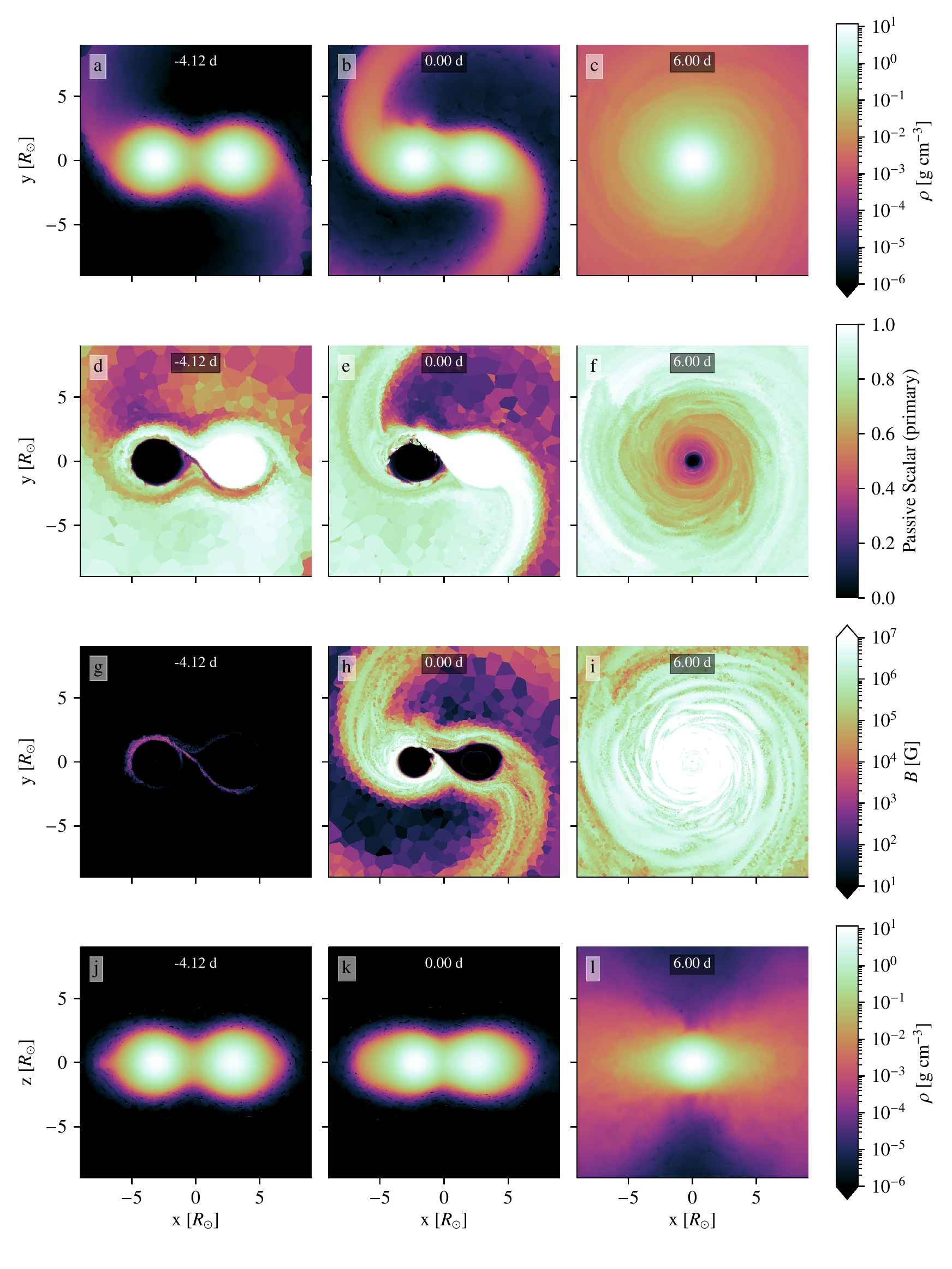}
\caption{\textbf{Evolution of density, magnetic-field strength and a passive scalar in the simulation of a merger of two main-sequence stars.}
Panels \textbf{a}--\textbf{c} show density snapshots in the orbital plane while panels \textbf{j}--\textbf{l} are edge-on views of the density. The passive scalar (white colour; panels \textbf{d}--\textbf{f}) indicates material from the $9\,\msun$ primary and thus visualises the mixing of the two progenitor stars during the merger. The passive scalar and the magnetic-field strengths (panels \textbf{g}--\textbf{i}) are shown in the orbital plane. The times given in each panel are relative to the time when the cores of the two star coalesce (middle panels \textbf{b}, \textbf{e}, \textbf{h}, and \textbf{k}).}
\label{fig:rho-b-evolution}
\end{figure}

%
%
Snapshots of the density, a passive scalar indicating material from the primary star and the absolute magnetic-field strength of the \threed MHD simulation are shown in Fig.~\ref{fig:rho-b-evolution}. Upon contact of the binary, a dynamical phase of mass transfer with rates as high as $17\,\msun\,\yr^{-1}$ sets in from the initially $9\,\msun$ primary star onto the initially $8\,\msun$ secondary star. Mass is lost through the outer Lagrangian points, draining angular momentum and thereby accelerating the coalescence. The accretion stream hits the surface of the secondary star and produces shear. It is in this accretion stream of size $0.8$ solar radii ($\rsun$) that the magnetic field is amplified on a e-folding timescale of about $0.2\text{--}1\,\day$ (Fig.~\ref{fig:rho-b-evolution}d and~\ref{fig:rho-b-evolution}g). The maximum magnetic-field strength saturates at about $10^8\,\gauss$, which corresponds to an amplification factor of about $10^6$. At saturation, the magnetic energy is comparable (about 5--30\%) to the turbulent energy which is the driving source of the magnetic-field amplification process. In the final merger, the amplified field is advected throughout the merger product and is finally also present in the core of the merger remnant. When the primary star is disrupted around the secondary and the cores of the two stars merge, vortices form at the interface of the former two cores (Fig.~\ref{fig:rho-b-evolution}e) that further contribute to the magnetic field amplification process (see also Supplementary Video S1). The maximum ratio of magnetic to gas pressure reaches 30\% in localised regions but is less than 1\% in the phase that leads up to the merger.

The local conditions in the differentially-rotating accretion stream (rotational frequency of $\Omega \approx 10\,\day^{-1}$, \alfven velocity of $\approx 1\,\kms$ and rotational shear $q=-\mathrm{d}\ln \Omega / \mathrm{d} \ln r \approx 0.4$) indicate that the magneto-rotational instability (MRI) \cite{1991ApJ...376..214B} is the key agent providing the turbulence to exponentially amplify the magnetic fields. In the shearing layer, the fastest growing mode of the MRI has a characteristic size of $0.1\,\rsun$ and growth timescale of $0.5\,\day$ \cite{2016MNRAS.456.3782R}, agreeing well with the size of the accretion stream and the observed growth timescale of the magnetic fields in our merger simulation.

Because of the large amount of angular momentum in the binary progenitor, a torus of about $3\,\msun$ forms that surrounds the central, spherically symmetric $14\,\msun$ core of the merger product (Fig.~\ref{fig:rho-b-evolution}\c and~l). The central merger remnant is in solid-body rotation while the centrifugally-supported torus rotates at sub-Keplerian velocities. The innermost core of the merger remnant consists of material from the former secondary star while the torus is dominated by core material of the former primary star (Fig.~\ref{fig:rho-b-evolution}f). The layers in between are a mixture of both progenitor stars.

We continue the \threed MHD simulation for $10\,\day$ after the actual merger, that is about $5\,\day$ after the merger remnant has settled into its final core-torus structure. This corresponds to roughly $5$ \alfven crossing timescales through the $14\,\msun$ core and we do not observe significant changes in the magnetic field structure and strength. The ratio of toroidal to total magnetic field energy is 80--85\% which is in a regime where magnetic-field configurations are thought to be stable in stellar interiors \cite{2006A&A...450.1077B}. Because of the high conductivity of stellar plasmas, Ohmic decay of the field might only occur on a timescale similar to or even longer than the stellar lifetime (see Supplementary Information). We therefore expect the magnetic field to be long-lived. 

%
%
The torus is expected to be accreted onto the central merger remnant on a timescale set by turbulent viscosity, that is within $0.02\text{--}0.1\,\yr$. Compared to this, cooling is slow ($100\text{--}1000\,\yr$) such that most of the torus will transform into an extended envelope rather than a thin disk. For the long-term evolution of the merger remnant we therefore assume that most mass, namely $16.9\,\msun$, ends up in the merger product and only less than $0.1\,\msun$ remains in a thin disk that carries most of the angular momentum (see Methods and Extended Data Fig.~\ref{fig:tvisc-tcool}).
 
Under these assumptions, we follow the evolution of the merger product with the \mesa stellar evolution code. As suggested by the \threed MHD simulations, we assume the formed remnant to rotate rigidly at a rate close to break-up. The magnetic flux at the end of our \threed simulation at a mass coordinate of $16.9\,\msun$ is $3.5\times10^{27}\,\gauss\,\mathrm{cm}^2$ (radius of $54\,\rsun$ and magnetic field of $80\,\gauss$) such that the surface magnetic-field strength of the merger remnant on the main sequence would be $9\,\mathrm{kG}$ for a radius of about $5\,\rsun$. This is well within observed surface field strengths of magnetic stars \cite{2009ARA&A..47..333D,2015A&A...582A..45F,2017MNRAS.465.2432G}. Because it is impossible to follow the evolution of an inherently \threed magnetic field in a \oned stellar evolution code, we assume that the radial magnetic-field strength in our \oned model follows that of a magnetic dipole. It contributes to internal angular-momentum transport and additional angular-momentum loss from the surface through a magnetised stellar wind (magnetic braking) but has otherwise no influence on the structure and evolution (see Methods for more details).

\begin{figure}
\centering
\includegraphics[width=1.0\textwidth]{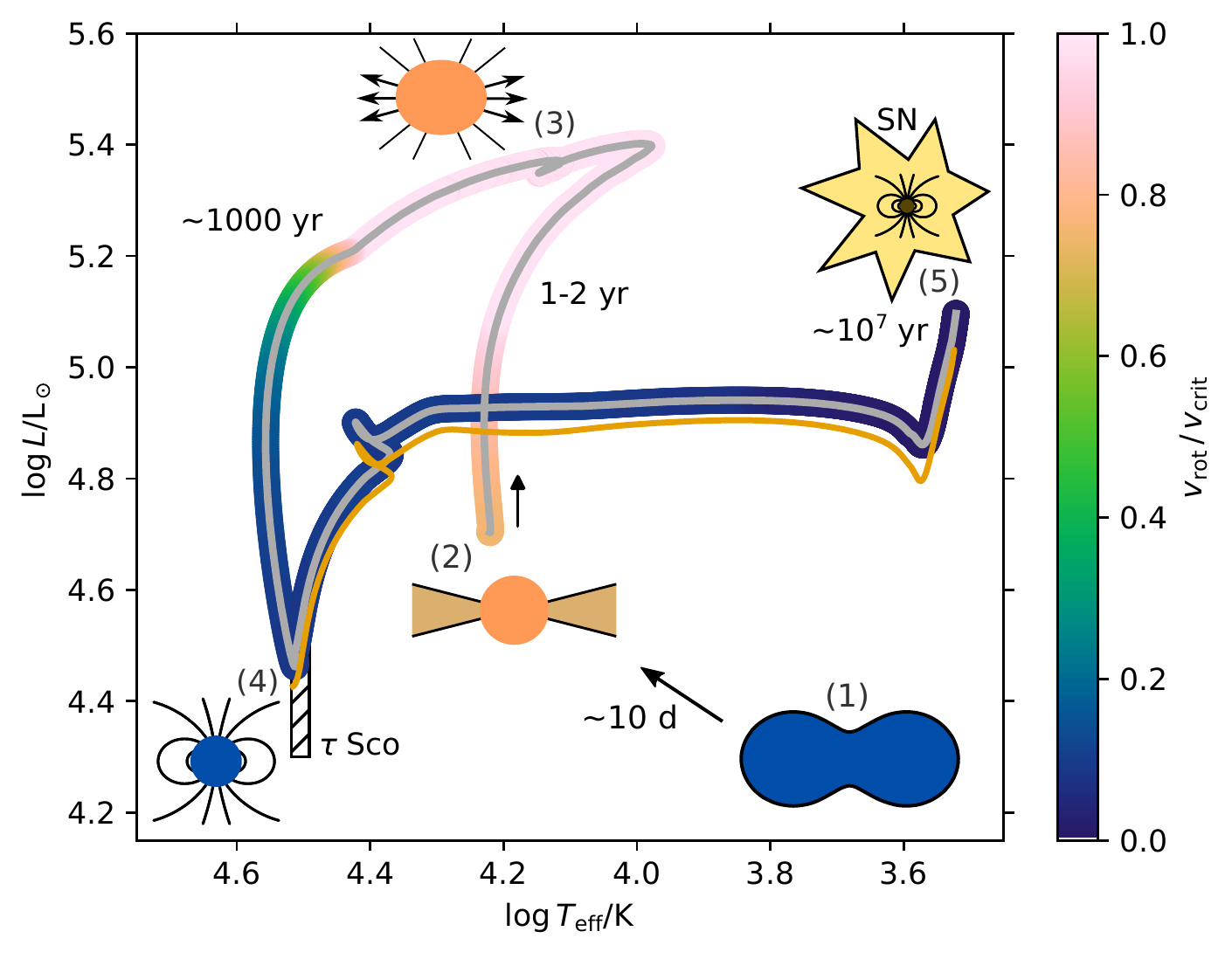}
\caption{\textbf{Long-term evolution of the merger product in the Hertzsprung--Russell diagram}.
After most of the torus is accreted, the merger remnant rotates rapidly and a thermal relaxation phase sets in during which the star first expands before it contracts back to the main-sequence (gray line). The direction of evolution is indicated by the black arrow at the beginning of this phase. The colour-coding shows the surface rotational velocity, $v_\mathrm{rot}$, in terms of the critical Keplerian velocity, $v_\mathrm{crit}$. Once on the main-sequence, the merger product continues its evolution similar to that of a genuine single star of the same mass of $16.9\,\msun$ (orange line). The hatched box indicates observations of \tausco ($T_\mathrm{eff}\, {=}\, 31,000\text{--}33,000\,\mathrm{K}$, $\log L/\lsun\, {=}\, 4.3\text{--}4.5$; see Table~\ref{tab:comp-param}). The small cartoons are artist's impressions of key evolutionary phases: (1) the contact phase before the actual merger, (2) the merger product with its torus, (3) during the thermal relaxation as a critically rotating star shedding mass, (4) as a main-sequence star with a strong surface magnetic field and (5) after the terminal supernova explosion that may form a magnetar.}
\label{fig:hrd}
\end{figure}

Because of the coalescence, the stellar interior is heated and the star is out of thermal equilibrium. A thermal relaxation phase sets in and, within $1\text{--}2\,\yr$, the star reaches a maximum radius of ${\approx}\,200\,\rsun$ and a maximum luminosity of $\log L/\lsun \approx 5.4$. After that, it contracts for a few $1000\,\yr$ towards the main-sequence from where on it continues its evolution close to that of a genuine single star of initially $16.9\,\msun$ (Fig.~\ref{fig:hrd}).

During the thermal expansion, the star reaches critical rotation at the surface, leading to additional mass and angular-momentum loss. Less than $0.01\,\msun$ are lost, taking away roughly 7\% of the star's total angular momentum. The magnetic fields keep the star close to solid-body rotation such that the stellar surface rotational velocity $v_\mathrm{rot}$ evolves according to
\begin{equation}
v_\mathrm{rot} = \frac{J}{r_\mathrm{g}^2 M R_{*}} \propto \left( r_\mathrm{g}^2 R_{*} \right)^{-1}.
\label{eq:vrot}
\end{equation}
Here, $J$ is the total angular momentum of the star, $M$ the mass, $R_{*}$ the radius and $r_\mathrm{g}$ the radius of gyration. During the subsequent contraction phase (radius decrease by a factor of ${\approx}\,4$), the surface spins down by a factor of ${\approx}\,5$ from critical rotation to ${\approx}\,50\,\kms$ (${\approx}10\%$ of critical rotation; see also Supplementary Information). The spin-down is not driven by angular-momentum loss but by an internal restructuring of the star that increases $r_\mathrm{g}^2$ by a factor of ${\approx}\,20$ and thus fully explains the spin-down (Eq.~\ref{eq:vrot}). In this case, magnetic braking is unimportant because the assumed surface magnetic field is weak ($<10\,\gauss$). The spin of the merger product on the main-sequence is thus set by the angular momentum that remains in the merger after the viscous accretion of the torus and the corresponding outward angular-momentum transport. However, the details of the viscous evolution of the torus and, in particular, the importance of magnetic braking, which depends on the evolution of the magnetic field after the merger, are currently uncertain.

Once back in thermal equilibrium, the merger product is a slow rotator with effective temperature, luminosity and surface gravity in excellent agreement with \tausco (Table~\ref{tab:comp-param} and Fig.~\ref{fig:hrd}), despite the fact that we have not fine-tuned the merger model (\eg by varying the initial mass and evolutionary state of the progenitor stars). Consequently, our merger product will also look like a rejuvenated blue straggler compared to other, apparently older stars in the Upper Scorpius association mainly because of the shorter lifetime associated with the now more massive star. \tausco is enriched in nitrogen on the surface, which is currently not reproduced by our model. However, on average, the envelope of our merger model is nitrogen-rich because it is made out of core material of the former primary star (Fig.~\ref{fig:rho-b-evolution}f). These enriched layers could easily be exposed by additional mass-loss or could be mixed to the stellar surface. For example, we have not considered mixing induced by the magnetic fields or during the viscous evolution of the torus. In conclusion, our merger scenario is able to explain the magnetic nature, atmospheric parameters, slow rotation and blue-straggler status of \tausco.

%
%
Strong amplification of magnetic fields is also observed in the coalescence of white dwarfs leading up to Type Ia supernovae \cite{2015ApJ...806L...1Z}, the merger of neutron stars producing gravitational waves and short-duration gamma-ray bursts \cite{2006Sci...312..719P}, and the common-envelope phase of a star spiralling into the envelope of a giant companion \cite{2016MNRAS.462L.121O}. Together with our simulations, it therefore seems that such dynamic and turbulent phases in the lives of stars provide the right conditions to produce strong magnetic fields. Consequently, also the coalescence of other main-sequence stars and stars in different evolutionary phases (\eg pre-MS) are expected to generate strong magnetic fields and therefore makes merging a plausible way to explain magnetic massive stars. 

%
%
The magnetic flux in the innermost $1.5\,\msun$ of our merger model at the end of the \threed simulation is about $4\times10^{28}\,\gauss\,\mathrm{cm}^2$ (radius of about $0.5\,\rsun$ and magnetic-field strength of $10^7\,\gauss$). If all of the magnetic flux is conserved until core collapse of the merger product, a resulting neutron star of $10\,\mathrm{km}$ radius would have a surface magnetic-field strength of about $10^{16}\,\gauss$. Hence, our model appears to be able to explain the strong magnetic fields inferred for magnetars ($10^{13}\text{--}10^{15}\,\gauss$) \cite{2014ApJS..212....6O}. The birthrate of magnetars in our Galaxy of about 0.3 per century \cite{2008MNRAS.391.2009K} and the rate of Galactic core-collapse supernovae of about 2 per century \cite{2006Natur.439...45D} suggest that 15\% of all Galactic core-collapse supernovae have produced a magnetar, which is consistent with the $10\%$ incidence of magnetic massive stars. Taken together, this makes our merger model a promising way to explain the strong magnetic fields observed in a subset of massive stars and also the origin of magnetars.

\clearpage

\clearpage

\clearpage
\begin{table}
\caption{\label{tab:comp-param}\textbf{Comparison of effective temperature ($T_\mathrm{eff}$), luminosity ($\log L/\lsun$) and surface gravity ($\log g$) of the merger product after thermal relaxation with observations of \tausco.}}
\centering
\begin{tabular}{lccc}
\toprule 
 & $T_\mathrm{eff}/\mathrm{K}$ & $\log L/\lsun$ & $\log g/\mathrm{cm}\,\mathrm{s}^{-2}$ \\
\midrule
\midrule
Merger model & ${\approx}\,32500$ & ${\approx}\,4.50$ & ${\approx}\,4.17$ \\
Mokiem et al.\ 2005 \cite{2005A+A...441..711M} & $31900^{+500}_{-800}$ & $4.39\pm0.09$ & $4.15^{+0.09}_{-0.14}$ \\
Sim{\'o}n-D{\'i}az et al.\ 2006 \cite{2006A+A...448..351S} & $32000\pm1000$ & $4.47\pm0.13$ & $4.00\pm0.10$ \\
Nieva and Przybilla 2014 \cite{2014A+A...566A...7N} & $32000\pm300$ & $4.33\pm0.05$ & $4.33\pm0.06$ \\
\bottomrule
\end{tabular}
\end{table}

%
%
\clearpage

\renewcommand*{\thefootnote}{\fnsymbol{footnote}}
\setcounter{footnote}{1}

\section*{Methods}
%
%
\subsection*{\threed MHD merger simulations}\label{sec:methods-arepo}

\paragraph*{\arepo's MHD solver}\label{sec:mhd-solver}
The \arepo code uses a second-order finite-volume method to solve the ideal MHD equations on an unstructured grid \cite{2010MNRAS.401..791S,2011MNRAS.418.1392P,2013MNRAS.432..176P}. The grid is generated in each timestep from a set of mesh-generating points that are allowed to move along with the flow, thus ensuring a nearly Lagrangian behaviour while regularising the mesh by adding an additional term. The fluxes over the cell boundaries are computed using the HLLD solver and the divergence of the magnetic field is effectively controlled employing the Powell scheme \cite{1999JCoPh.154..284P}, as shown in refs.\citenum{2011MNRAS.418.1392P,2013MNRAS.432..176P}. We use ideal MHD here because the resistivity is very small in the highly conducting plasma of stellar interiors (see also Supplementary Information).

\paragraph*{Initialisation of the binary progenitor}\label{sec:arepo-binary-setup}
The binary progenitors have an initial helium mass fraction of $Y=0.2703$ and solar metallicity $Z=0.0142$ \cite{2009ARA&A..47..481A}. The stellar structures are imported from \oned \mesa \cite{2011ApJS..192....3P,2013ApJS..208....4P,2015ApJS..220...15P,2018ApJS..234...34P} models (version 9793) that employ exponential convective-core overshooting with a parameter of $f_\mathrm{ov}=0.019$, which effectively corresponds to a step convective-core overshooting of about $0.16$ pressure scale heights.

Mapping the \oned stellar structures into a \threed hydrodynamics code leads to discretisation errors in the hydrostatic equilibrium; thus, relaxation methods \cite{2017A&A...599A...5O} are employed to create stable stellar models. The \oned stellar models from \mesa are mapped onto an unstructured grid consisting of HEALPix distributions on spherical shells \cite{2005ApJ...622..759G}. These models are relaxed in \arepo with a damping scheme to remove spurious velocities due to discretisation errors. Both stellar models have been processed this way and the resulting stars are stable according to the criteria discussed in ref.\citenum{2017A&A...599A...5O}. The initial seed magnetic field was set up in a dipole configuration with a polar surface field strength of $1\,\gauss$.

The relaxed single-star models are subsequently used to set up the binary star merger. Although one would like to follow the merger beginning from Roche-lobe overflow until the actual merger, this process takes too long for the simulation to be computationally feasible. Hence, we speed up the merging process by artificially decelerating each cell for a certain time (about 1.5 orbits) and starting the actual calculation from there. The duration of this deceleration phase influences the outcome of the merger only marginally (see Supplementary Information).

%
%
\subsection*{\oned long-term evolution of the merger product}\label{sec:methods-mesa}

The \threed MHD merger simulations have produced strong magnetic fields that are relevant for the further evolution of the merger product. Although the generated magnetic fields are too weak to directly affect the stellar structure, they can contribute to the angular-momentum transport through the stellar interior and may lead to additional angular-momentum loss from the stellar surface through magnetised winds (magnetic braking). In this way, the magnetic fields influence the evolution of the star. In the following, we describe the assumed magnetic field structure in our \oned stellar evolution models, and our implementation of the interior angular-momentum transport through the magnetic fields and magnetic braking. We then explain how our \oned models are set-up based on the outcome of the \threed MHD simulations. For the \oned computations, we use the \mesa stellar evolution code in version 9793 \cite{2011ApJS..192....3P,2013ApJS..208....4P,2015ApJS..220...15P,2018ApJS..234...34P}.

\paragraph*{Assumed large-scale magnetic field in \oned computations}\label{sec:mesa-b-field}
As mentioned in the main text, it is not possible to follow the evolution of a \threed magnetic field in a \oned stellar evolution code. Moreover, the final configuration of the magnetic field after the accretion of the torus is still uncertain. Hence, we assume that the radial magnetic-field strength in our \oned model follows that of a magnetic dipole, $B(r) = \mu_\mathrm{B} r^{-3}$, with dipole moment $\mu_\mathrm{B} = 2\times10^{37}\,\mathrm{G}\,\mathrm{cm}^3$. This assumption is conservative in the sense that it results in a surface magnetic field of the merger product on the main-sequence of a few hundred Gauss which is lower than that expected from magnetic flux freezing of our \threed model. Using larger or smaller magnetic-dipole moments does not affect our conclusions.

For a dipole magnetic field, the ratio of the field strength at the pole ($B_\mathrm{p}$) and the equator ($B_\mathrm{eq}$) is $B_\mathrm{p}/B_\mathrm{eq}=2$. The dipole field diverges for $r\rightarrow 0$ and we therefore cap its field strength to $10^9\,\gauss$. In our models, the applied magnetic dipole moment is reminiscent of that of \tausco, \ie a polar field strength of ${\approx}\,500\,\mathrm{G}$ for a ${\approx}\,5\,\rsun$ star \cite{2006MNRAS.370..629D}.

We further assume that the magnetic field is expelled from convective regions if the convective energy density $u_{\mathrm{conv}}$ is larger than the magnetic energy density $u_{\mathrm{B}}$, \ie if
\begin{equation}
u_{\mathrm{conv}}=\frac{1}{2}\rho v_{\mathrm{conv}}^{2} > u_{B}=\frac{B^{2}}{8\pi}.
\label{eq:b-field-conv}
\end{equation}
Here, $\rho$ is the gas density and $v_{\mathrm{conv}}$ the velocity of convective eddies as predicted by mixing-length theory. This treatment of the static magnetic field means that it only contributes to the angular-momentum transport in radiative regions and does not provide an efficient coupling of the convective core and radiative envelope in our models.

\paragraph*{Angular-momentum transport in the stellar interior through a large-scale magnetic field}\label{sec:mesa-ang-mom-transport}
We treat the transport of angular momentum through the stellar interior as a diffusive process. Magnetic fields cause Maxwell stresses and can thus transport angular momentum. To obtain the effective diffusion coefficient of this process (which we call effective viscosity $\nu_\mathrm{eff}$), we consider differentially-rotating, spherical shells and assume that the stresses due to magnetic fields are effectively similar to the classical Newtonian dynamic shear $S$,
\begin{equation}
S=\frac{\mathrm{d}F}{\mathrm{d}A}=\nu_{\mathrm{eff}}\rho\frac{\partial v}{\partial r},
\label{eq:newton-shear}
\end{equation}
where $\mathrm{d}F$ is the force exerted by the shear on an area $\mathrm{d}A$ and $\partial v/\partial r$ the radial gradient of the velocity $v$. In spherical coordinates ($r$, $\varphi$, $\theta$), the torque $\mathrm{d}\tau$ on a surface element $\mathrm{d}A=r^{2}\sin\theta\,\mathrm{d}\varphi\,\mathrm{d}\theta$ due to a shear force $\mathrm{d}F$ is given by
\begin{equation}
\mathrm{d}\tau = r \sin\theta \, \mathrm{d}F = r \sin\theta \, \nu_{\mathrm{eff}}\rho\frac{\partial v}{\partial r}\mathrm{d}A.
\label{eq:delta-torque}
\end{equation}
Introducing the angular velocity $\Omega$ ($v=r\sin\theta\Omega$), we have $\partial v/\partial \Omega = r\sin\theta$ and thus $\partial v/\partial r = r\sin\theta \partial\Omega/\partial r$. Integrating Eq.~\ref{eq:delta-torque} over $\varphi$ and $\theta$, we obtain the overall torque on a shell at radius $r$ as
\begin{equation}
\tau = \frac{8\pi}{3}\nu_{\mathrm{eff}}r^{4}\rho\frac{\partial\Omega}{\partial r}.
\label{eq:torque}
\end{equation}
From a physical point of view, the shear exerted by magnetic fields will reduce differential rotation and attempts to establish solid-body rotation, \ie $\partial\Omega/\partial r=0$. The amount of angular momentum $\Delta J$ that needs to be transported to achieve solid-body rotation in neighbouring, differentially rotating shells is $\Delta J=I\Delta\Omega$, where $I$ is the moment of inertia. The angular-momentum transport across a shell of thickness $\Delta r$ occurs with an \alfven velocity $v_{\mathrm{A}}=B/\sqrt{4\pi\rho}$, \ie on an \alfven timescale of $\tau_{\mathrm{A}}=\Delta r/v_\mathrm{A}$, such that 
\begin{equation}
\frac{\mathrm{d}J}{\mathrm{d}t} \approx \frac{\Delta J}{\tau_{\mathrm{A}}} = \frac{I\left(\partial\Omega/\partial r\right)\Delta r}{\tau_{\mathrm{A}}} = I\left(\frac{\partial\Omega}{\partial r}\right)v_{A}.
\label{eq:torque2}
\end{equation}
Equating Eq.~(\ref{eq:torque2}) and Eq.~(\ref{eq:torque}), we find the desired effective viscosity for angular-momentum transport in differentially rotating shells because of a large-scale magnetic field,
\begin{equation}
\nu_{\mathrm{eff}} = \frac{3I}{8\pi r^{4}\rho}v_{\mathrm{A}}.
\label{eq:nu-eff}
\end{equation}
The moment of inertia of a single shell in a stellar evolution model depends on the spatial discretisation. To make the effective diffusion coefficient resolution-independent, we define `shells' to have a thickness of 20\% of the local pressure scale height $H_\mathrm{P}$.

For thin shells of mass $\Delta m$, the moment of inertia is $I=2/3 \Delta m r^{2}$. The effective magnetic viscosity from Eq.~(\ref{eq:nu-eff}) is then approximately $\nu_{\mathrm{eff}} = \Delta r v_{\mathrm{A}}$ with $\Delta r=0.2 H_\mathrm{P}$ for our definition of a `shell'. This is similar to the general form of a diffusion coefficient, $D \propto l v$, for a diffusion process over a length scale $l$ with characteristic velocity $v$. We modulate the effective viscosity with a factor $f_{\mathrm{A}}$ that is thought to adjust the timescale over which solid-body rotation is achieved in neighbouring shells. We set $f_{\mathrm{A}}=0.5$ in our calculations and note that small variations in $f_{\mathrm{A}}$ hardly change our results. Taken together with our choice of $f_{\mathrm{A}}$, the effective magnetic viscosity is equivalent to $\nu_\mathrm{eff} \approx 0.1 H_\mathrm{P} v_\mathrm{A}$.

In the above analysis, we have not made explicit assumptions on the magnetic-field geometry, but it will of course matter in reality. For example, if there is no radial B-field component, the Maxwell stress is zero such that there is no angular-momentum transport in the radial direction through the magnetic field. In our approach, the field geometry enters indirectly through the \alfven velocity which depends on the absolute magnetic-field strength that itself is a function of radius $r$. Furthermore, we made the assumptions that the magnetic field is able to establish solid-body rotation in neighbouring shells and that the stresses are similar to classical dynamic shear stresses.

\paragraph*{Magnetic braking}\label{sec:mesa-magnetic-braking}
Stellar winds can couple to large-scale magnetic fields and thereby enhance the loss of angular momentum, a process called magnetic braking. In a simple explanation one can imagine that the magnetic field establishes solid-body rotation in the out-flowing wind with the stellar surface out to the Alfv\'en radius, $R_{\mathrm{A}}$. In this way, the wind takes away the specific angular momentum from the Alfv\'en radius instead of the stellar surface. To be more precise, the angular-momentum loss from the stellar surface because of magnetic braking, $\mathrm{d}J_{\mathrm{mb}}$, is due to Poynting stresses caused by bent magnetic-field lines \cite{2009MNRAS.392.1022U} such that the resulting torque is
\begin{equation}
\frac{\mathrm{d}J_{\mathrm{mb}}}{\mathrm{d}t} = \frac{2}{3} \dot{M} \Omega_{*} R_{\mathrm{A}}^{2}.
\label{eq:magnetic-braking}
\end{equation}
In this equation, $\dot{M}$ is the stellar wind mass loss rate, $\Omega_{*}$ is the stellar surface angular velocity and the factor $2/3$ accounts for the momentum of inertia of thin spherical shells.

In MHD simulations of magnetic braking of hot, massive stars, the Alfv\'en radius in Eq.~\ref{eq:magnetic-braking} is found to be 
\begin{equation}
\frac{R_{\mathrm{A}}}{R_{*}} \approx 0.29+\left(\eta_{*}+0.25\right)^{1/4}
\label{eq:alfven-radius-mb}
\end{equation}
with $R_{*}$ the stellar radius, $\eta_{*}$ the wind magnetic confinement parameter,
\begin{equation}
\eta_{*}=\frac{B_{\mathrm{eq}}^{2}R_{*}^{2}}{\dot{M}v_{\infty}},
\end{equation}
$B_{\mathrm{eq}}$ the equatorial, surface magnetic-field strength and $v_{\infty}$ the terminal wind velocity \cite{2009MNRAS.392.1022U}. 

For the terminal wind velocity, we use the observational results for O to F stars \cite{1995ApJ...455..269L},
\begin{equation}
\frac{v_{\infty}}{v_{\mathrm{esc}}}=\begin{cases}
0.7 & \text{for }\log T_{\mathrm{eff}}\leq4.0,\\
1.3 & \text{for }4.0<\log T_{\mathrm{eff}}\leq4.32,\\
2.6 & \text{for }4.32<\log T_{\mathrm{eff}}.
\end{cases}
\end{equation}
Here, the escape velocity is defined as
\begin{equation}
v_{\mathrm{esc}}=\sqrt{\frac{2GM(1-\Gamma_{\mathrm{es}})}{R_{*}}}
\end{equation}
with $\Gamma_{\mathrm{es}}$ the electron-scattering Eddington factor and $M$ the stellar mass.

From a technical point of view, stellar winds in stellar evolution codes take away the specific angular momentum of their former Lagrangian mass shells. In our models, the additional angular momentum lost through magnetic braking is then taken away from a thin surface layer after the mass shells that are lost in the wind have been removed.

\paragraph*{Import of the merger remnant into a \oned stellar evolution code}\label{sec:mesa-import}
Right after the merger, the evolution is mainly driven by that of the torus and its interplay with the central star. Two timescales are most relevant: the accretion and cooling timescales ($\tau_\mathrm{acc}$ and $\tau_\mathrm{cool}$, respectively). The accretion timescale sets the time over which the torus is accreted by the central remnant while the cooling timescale describes the time over which the torus loses the heat produced by the accretion.

Matter in the rotationally-supported torus can only be accreted onto the central star if its angular momentum is transported outwards; hence, the accretion timescale is given by the angular-momentum transport timescale. We assume that the matter and angular momentum flow in the torus can be described by an $\alpha$-disk model with an effective viscosity\footnote{Here and throughout, the word ``viscous'' is used in the phenomenological sense of an effective viscosity which acts on large scales due the presence of an enhanced turbulent transport. It should not be confused with the true microscopic particle viscosity, which is negligible for the problems of interest here.} $\alpha$ that, for example, might be provided by the magnetic fields or the magneto-rotational instability \cite{1991ApJ...376..214B}. Using the mass accretion rate for such a disk model,
\begin{equation}
\dot{M}_{\mathrm{acc}} \approx 3 \alpha \left(\frac{h}{r}\right)^2 \Omega M_\mathrm{disk},
\label{eq:disk-mdot}
\end{equation}
the accretion timescale of the torus is
\begin{equation}
\tau_{\mathrm{acc}} = \frac{M_{\mathrm{disk}}}{\dot{M}_{\mathrm{acc}}} = \frac{1}{3}\frac{r^{2}}{\alpha h^{2}\Omega} = 0.02\,\yr\,\left(\frac{10^{-2}}{\alpha}\right)\left(\frac{r/h}{2}\right)^{2}\left(\frac{\mathrm{h}^{-1}}{\Omega}\right).
\label{eq:tau-acc}
\end{equation}
Here, $h/r$ is the ratio of disk height and radius, $M_\mathrm{disk}$ the mass in the disk and $\Omega$ the angular velocity of the disk which generally depends on radius. In our case, the accretion timescale is equivalent to the viscous timescale $\tau_\mathrm{visc}$.

Mass accretion leads to (turbulent) heating through the release of gravitational potential energy, $E_\mathrm{grav}$. On the one hand, if this energy can be lost efficiently from the system via fast cooling ($\tau_{\mathrm{cool}}\ll\tau_{\mathrm{acc}}$), the torus becomes thinner or at least keeps its shape. On the other hand, if the cooling is inefficient ($\tau_{\mathrm{cool}}\gg\tau_{\mathrm{acc}}$), the torus becomes thicker and evolves into a thermally supported extended envelope. Assuming that the star-torus structure radiates at a fraction $f_\mathrm{Edd}$ of its Eddington luminosity, $L_{\mathrm{Edd}}$, and that photon cooling is the dominant cooling process, the cooling timescale can be approximated as
\begin{align}
\tau_{\mathrm{cool}} &= \frac{E_{\mathrm{grav}}}{f_\mathrm{Edd} L_{\mathrm{Edd}}} = \frac{1}{f_\mathrm{Edd}} \frac{GM_{\mathrm{core}}M_{\mathrm{disk}}/R_{\mathrm{core}}}{4\pi G(M_{\mathrm{core}}+M_{\mathrm{disk}})c/\kappa} \nonumber \\
 & \approx  0.8\times10^{3}\,\yr\,\left(\frac{1}{f_\mathrm{Edd}}\right) \left(\frac{1+X}{1.7}\right)\left(\frac{M_{\mathrm{core}}M_{\mathrm{disk}}/\mathrm{M}_{\odot}}{M_{\mathrm{core}}+M_{\mathrm{disk}}}\right)\left(\frac{R_{\odot}}{R_{\mathrm{core}}}\right),
\label{eq:tau-cool}
\end{align}
where $M_\mathrm{core}$ and $R_\mathrm{core}$ are the mass and radius of the central star, and $\kappa$ is the opacity. In the last step, we assumed that the opacity is dominated by electron scattering, \ie $\kappa=0.2(1+X)\,\mathrm{cm}^{2}\,\mathrm{g}^{-1}$ with the hydrogen mass fraction $X$. Even for $f_\mathrm{Edd}=1$, the cooling time in our case is of order $500\text{--}700\,\yr$ ($M_{\mathrm{core}}=14\,\mathrm{M}_{\odot}$, $M_{\mathrm{disk}}=3\,\mathrm{M}_{\odot}$, $R_{\mathrm{core}}=3\text{--}4\,\mathrm{R}_{\odot}$) and thus significantly longer than the accretion timescale in Eq.~(\ref{eq:tau-acc}). The expectation therefore is that the torus is rapidly accreted onto the central star and evolves into a thermally supported, extended envelope before its thermal relaxation and cooling process sets in.

These arguments are analogous to previous work on the merger remnant of two white dwarfs \cite{2012ApJ...748...35S}. More detailed simulations of the viscous evolution of this double white-dwarf merger remnant \cite{2012MNRAS.427..190S} support the (analytic) expectations \cite{2012ApJ...748...35S} and indeed show a rapid transformation of the torus into a thermally supported envelope. Given the similarity of the physical situation and the accretion and cooling timescales in our case, it therefore seems reasonable that large fractions of our torus will also evolve into a thermally supported envelope on a viscous timescale.

The accretion and cooling timescales (Eqs.~\ref{eq:tau-acc} and~\ref{eq:tau-cool}) depend on radius through the radially-declining angular velocity $\Omega$ of the torus and the radius $R_\mathrm{core}$ at which matter is accreted onto the central star. At a radius of about $54\,\rsun$, both timescales are comparable such that cooling is inefficient inside and efficient further out (Extended Data Fig.~\ref{fig:tvisc-tcool}). In our standard model, we therefore assume that the mass interior of $54\,\rsun$, \ie the innermost $16.9\,\msun$, transform into a star with an extended envelope on a viscous timescale. The remaining outer part of the torus is assumed to cool efficiently and evolve into a thin disk. This configuration then forms the initial condition of our \oned stellar evolution computations.

In order to import the outcome of the \threed simulation into the \oned stellar evolution code \mesa, we model a star that has the same chemical and thermal structure as the \threed merger remnant. We first relax a star of given total mass to the chemical structure of the \threed merger remnant before imposing the thermal structure by matching the \threed entropy profile. A comparison of the chemical and entropy structure of the \oned $16.9\,\msun$ merger remnant with the \threed profiles is shown in Extended Data Fig.~\ref{fig:3d-vs-1d}. Our \oned model closely matches the structure of the merger remnant of the \threed simulation.

Setting the rotational profile of the merger remnant requires further consideration. We argued above that the fast viscous evolution of the star-torus structure converts most of the torus into an extended envelope by transporting angular momentum outwards. This angular-momentum transport sets the initial conditions for our \oned merger evolution. In the viscous evolution of the remnant of a double white-dwarf merger, efficient outward angular-momentum transport is found such that the rotational profile of the central star remains a solid-body rotator and smoothly transitions into a near-Keplerian profile at the boundary between the central star and outer disk \cite{2012MNRAS.427..190S}. The same evolution and outcome is found by other authors who studied the aftermath of double white dwarf mergers within a prescribed viscosity model but also within more self-consistent MHD simulations \cite{2012ApJ...748...35S,2013ApJ...773..136J}. In all cases, a significant fraction of angular momentum is transported outwards allowing for the rapid accretion of a large fraction of the mass of the torus.

Also our \threed merger simulation shows that the central star reaches a state of solid-body rotation with the angular velocity matching that of the layer between star and torus, which is approximately 80\% of the Keplerian value. The surface of the central star does not reach 100\% Keplerian rotation because the torus is not only centrifugally supported but also thermally supported. Following these ideas, we assume that our merger remnant is a solid-body rotator that rotates at 90\% of the critical Keplerian velocity at the surface after the viscous evolution.

%
%
\renewcommand{\refname}{\vspace{-0.5cm}}
\bibliographystyle{naturemag}

%
%
\clearpage
\paragraph*{Supplementary Information}
is linked to the online version of the paper at www.nature.com/nature.

\paragraph*{Acknowledgements}
This work was supported by the Oxford Hintze Centre for Astrophysical Surveys which is funded through generous support from the Hintze Family Charitable Foundation. 
STO, FKR and FRNS acknowledge funding from the Klaus Tschira foundation.

\paragraph*{Author contributions}
F.R.N.S. initiated the project and carried out the \oned \mesa computations. S.T.O. carried out the \threed \arepo simulations. F.R.N.S. and S.T.O. mainly wrote the manuscript. Ph.P. and F.K.R. assisted with the \oned and \threed computations, respectively. S.A.B. in particular helped to analyse and understand the magnetic-field amplification process. R.P. and V.S. wrote the \arepo code and supported S.T.O. with the \threed simulations. All authors contributed to the analysis, discussion and writing of the paper.

\paragraph*{Competing interests}
The authors declare no competing interests.

\paragraph*{Additional information}\mbox{}
\newline
\textbf{Reprints and permissions information} is available at http://www.nature.com/reprints.
\newline
\textbf{Correspondence and requests for materials} should be addressed to F.R.N.S. and S.T.O.

%
%
\captionsetup[figure]{name={Extended Data Figure}}
\captionsetup[table]{name={Extended Data Table}}

\setcounter{figure}{0}
\setcounter{table}{0}

\clearpage
\begin{figure}
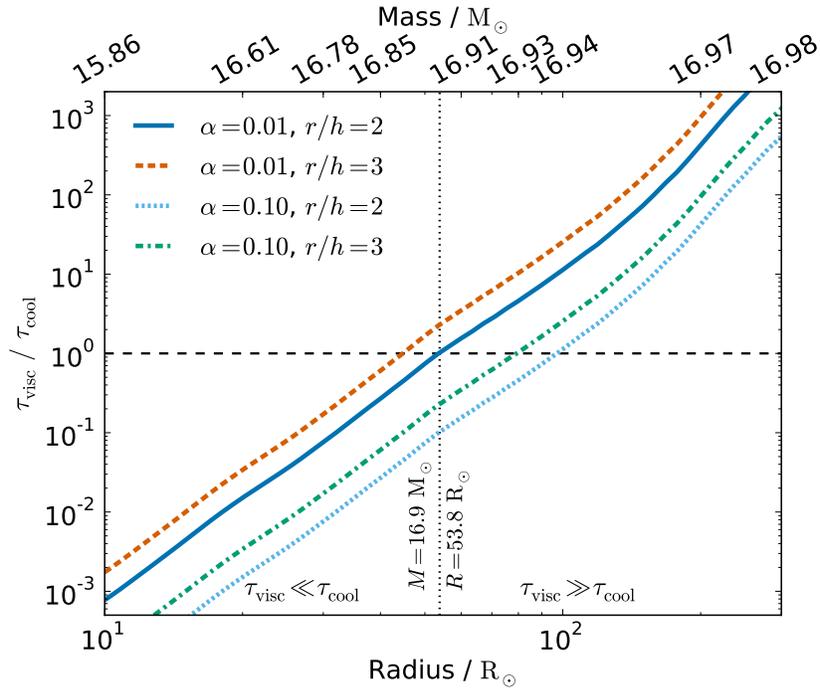

\begin{centering}
\includegraphics[width=0.7\textwidth]{{{tvisc-tcool-plot}}}
\par\end{centering}
\caption{\textbf{Ratio of viscous and cooling timescales.}
The ratio is shown as a function of radius (and mass) after $6\,\day$ of the merger for different disk thicknesses $h$ and viscosity parameters $\alpha$.}
\label{fig:tvisc-tcool}
\end{figure}

\begin{figure}
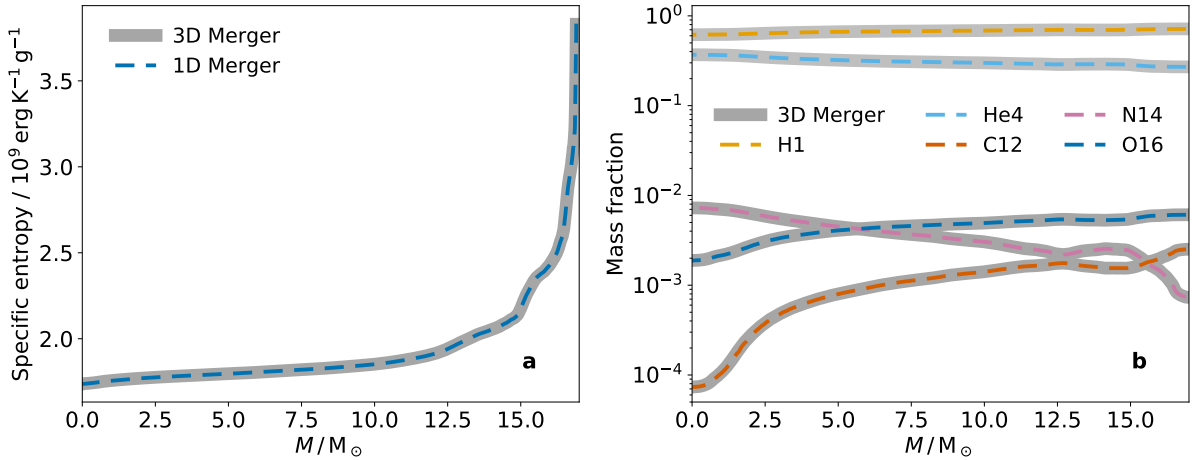

\begin{centering}
\includegraphics[width=1.0\textwidth]{{{init-condition-comparison-M-16.90}}}
\par\end{centering}
\caption{\textbf{Comparison of the final \threed and initial \oned merger product.}
Compared are the entropy (panel \textbf{a}) and hydrogen (H1), helium (He4), carbon (C12), nitrogen (N14) and oxygen (O16) mass fractions (panel \textbf{b}) of the \threed merger remnant (thick grey lines) and the \oned stellar model (dashed line).}
\label{fig:3d-vs-1d}
\end{figure}

%
%
\clearpage
\baselineskip12pt 

\captionsetup[figure]{name={Supplementary Figure}}
\captionsetup[table]{name={Supplementary Table}}

\renewcommand{\thefigure}{S\arabic{figure}}
\renewcommand{\thetable}{S\arabic{table}}
\renewcommand{\thesection}{S\arabic{section}}
\renewcommand{\theequation}{S.\arabic{equation}}

\setcounter{page}{1}
\setcounter{figure}{0}
\setcounter{table}{0}



\begin{center}
\LARGE{Supplementary Information for}
\end{center}

\begin{center}
\LARGE{\thetitle}
\end{center}

\begin{center}
\theauthor
\end{center}

\vspace{0.5cm}

\setlength{\parindent}{4ex}
\setlength{\parskip}{1.2mm}

\section*{Supplementary Text}

Here, we provide more details on a resolution and initial-condition study of the \threed MHD simulations (Sect.~\ref{sec:arepo-init-cond-study}), the magnetic-field saturation (Sect.~\ref{sec:b-field-saturation}), the Ohmic dissipation of magnetic fields (Sect.~\ref{sec:b-field-stability}), the internal restructuring of the merger model during the thermal relaxation that leads to its spin-down (Sect.~\ref{sec:restruct-star}) and a movie to assist the understanding of the magnetic-field amplification process in the merger (Sect.~\ref{sec:movies}).

\section{Resolution study and initial conditions}\label{sec:arepo-init-cond-study}

\begin{figure}
\begin{centering}
\includegraphics[width=0.75\textwidth]{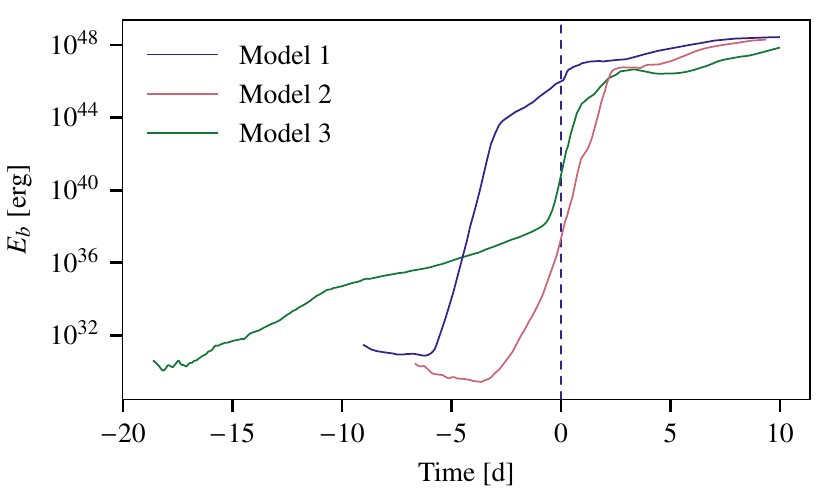}
\par\end{centering}
\caption{\textbf{Evolution of total magnetic field energy for different simulation setups.}
Model~1 is the standard run shown in the main text. Models~2 and~3 have a lower resolution.
Model~3 started with a larger initial separation. The times for all models are
normalized with the time of merger set to 0.}
\label{fig:magnetic-field-evolution}
\end{figure}

We ran simulations for different resolutions and initial binary setups to ensure that the amplification of the magnetic field is robust against variations of these parameters. The evolution of the total magnetic field energy over time is shown in Fig.~\ref{fig:magnetic-field-evolution}. The standard run shown in the main text is Model~1. The evolution of the magnetic energy is slightly different for the different configurations but the overall behaviour and the final energy are very similar. Model~2 tests a lower resolution for otherwise identical initial conditions. Model~3 was started at an earlier time with a larger initial separation. The resolution was set up with roughly $4\times 10^6$ cells for Model~1 and about $4\times 10^5$ cells for the other models.

\section{Magnetic-field saturation}\label{sec:b-field-saturation}

The magnetic field amplification switches off if the necessary conditions of that physical mechanism that drives the amplification process are no longer met. In case of the magneto-rotational instability (MRI) \cite{1991ApJ...376..214B}, this could be the case if the magnetic field becomes so strong that the fastest growing mode exceeds the spatial region of interest (\eg it becomes larger than the star), if the amplification timescale becomes excessively long or if there is no longer differential rotation. Such a situation may go along with an equipartition of the magnetic energy and that energy source (\eg differential or turbulent energy) that drives the magnetic-field amplification.

In our models, the initially fast, exponential magnetic-field amplification is consistent with being driven by the MRI. After the merger, the central star is in solid-body rotation such that the MRI cannot operate in this part any more. In the torus, however, the MRI is still active and the magnetic-field strength is indeed found to increase until the end of the simulation. The fastest growing MRI mode always fits into the central merger remnant and the MRI amplification timescale stays short compared to the runtime of our simulation.

\begin{figure}
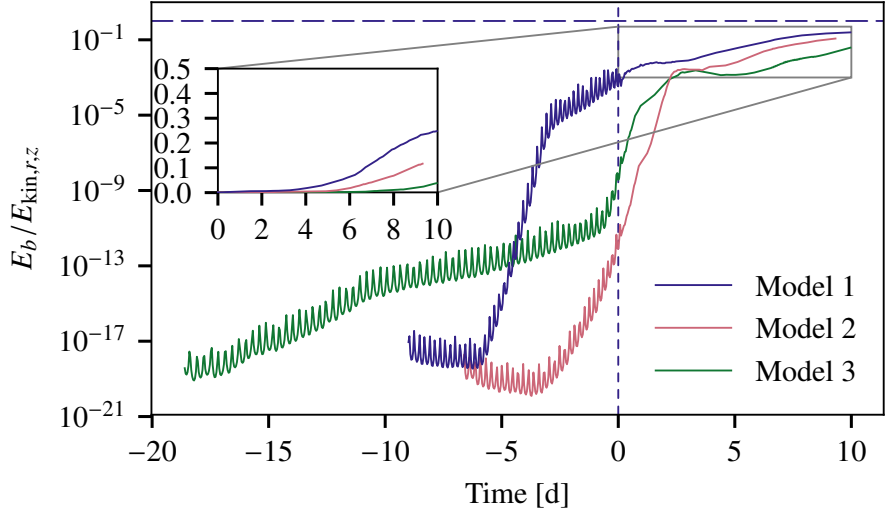

\begin{centering}
\includegraphics[width=0.75\textwidth]{{{magnetic-field-saturation}}}
\par\end{centering}
\caption{\textbf{Ratio of magnetic and radial, kinetic (\ie turbulent) energy in our \threed MHD simulations.}
The magnetic energies, $E_\mathrm{b}$, of our models approach equipartition with the turbulent energy, for which we use the kinetic energy of motions in radial and $z$ directions, $E_{\mathrm{kin},r,z}$, as proxy. The small inset shows the ratio of magnetic and kinetic energy on a linear scale and shows that our models reach values of about 5\%--30\%. The dashed horizontal line indicates equipartition of magnetic and kinetic energy. The small, periodic wiggles in the curves before coalescence are due to the orbital motion of the binary.}
\label{fig:b-field-saturation}
\end{figure}

Using the kinetic energy in radial and $z$ directions as proxy for the turbulent energy that is generally thought to drive the magnetic-field amplification, we find that the magnetic energy reaches a level of about 5\%--30\% of the kinetic energy in our models (Fig.~\ref{fig:b-field-saturation}). This supports the idea that the magnetic-field amplification ceases when approaching equipartition with the turbulent energy in the merger remnant.

\section{Ohmic dissipation of magnetic fields}\label{sec:b-field-stability}

Stable magnetic fields can diffuse out of the stellar interior by Ohmic resistivity and thereby dissipate \cite{2006A&A...450.1077B}. However, because the hot stellar interior is highly conducting, the resistivity is low and the dissipation of magnetic fields operates on a timescale that is longer or similar to the stellar lifetime. Indeed, for Spitzer's resistivity \cite{1962pfig.book.....S},
\begin{equation}
\eta = 7\times10^{11}\, \ln \Lambda\, \left(\frac{T}{\mathrm{K}}\right)^{-3/2}\, \mathrm{cm}^2\,\second^{-1}
\label{eq:spitzer-eta}
\end{equation}
with $T$ the temperature and $\ln \Lambda$ the Coulomb logarithm, which is of order 10 for stellar interiors, the diffusion timescale of magnetic fields is
\begin{equation}
\tau_\mathrm{diff} = R^2/\eta \approx 10^{8}\text{--}10^{11}\,\yr
\label{eq:tau-diff}
\end{equation}
for temperatures of $10^{5}\text{--}10^{7}\,\mathrm{K}$ and a length scale of $R=1\,\rsun$. These estimates depend on the still uncertain resistivity in stellar interiors but it appears that Ohmic dissipation of the amplified magnetic fields does not play a role in our case because the lifetime of the merger product is rather of order $10^7\,\yr$. It might however be relevant for some evolved stars \cite{2006A&A...450.1077B}.

\section{Restructuring of the stellar interior during the thermal relaxation phase after the merger}\label{sec:restruct-star}

\begin{figure}
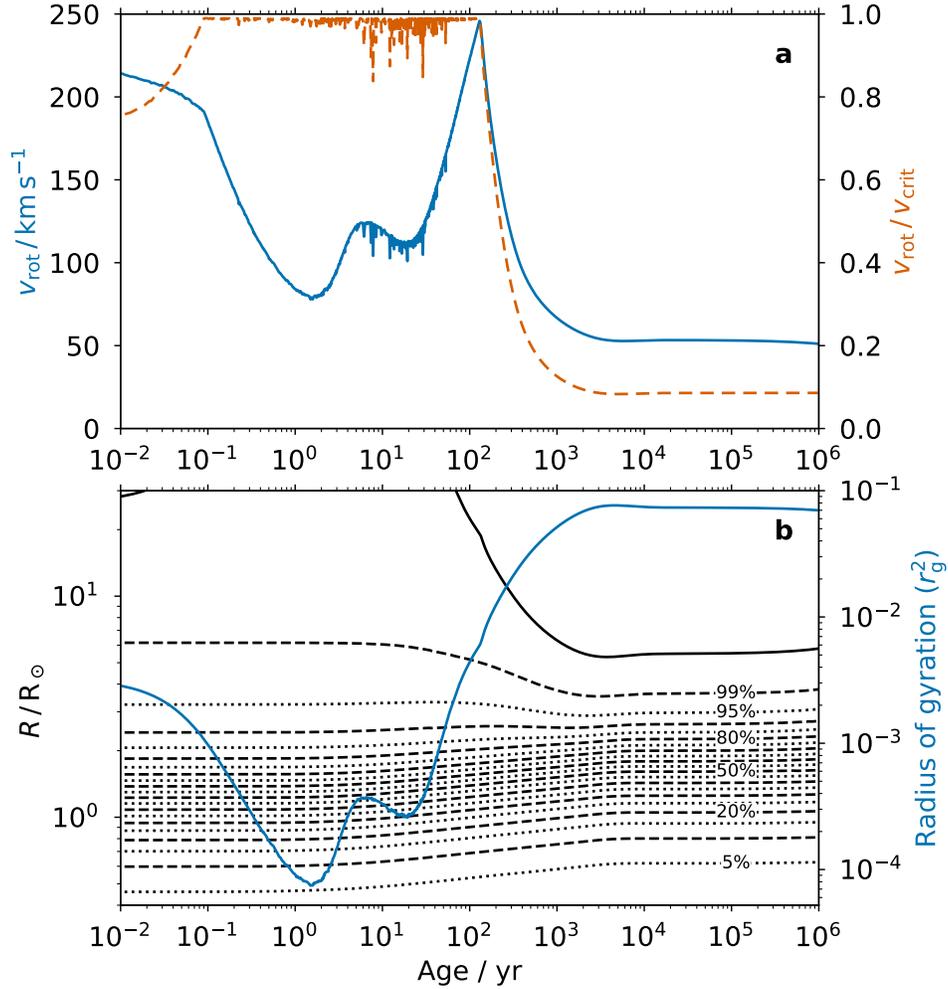

\begin{centering}
\includegraphics[width=0.8\textwidth]{{{internal-restruct}}}
\par\end{centering}
\caption{\textbf{Rotational velocity and internal mass readjustment of the \oned merger model.}
(\textbf{a}) Equatorial surface rotational velocity $v_\mathrm{rot}$ (blue solid line) and rotational velocity in terms of critical Keplerian velocity $v_\mathrm{crit}$ (red dashed line) as a function of time after the merger. (\textbf{b}) Radial location of various mass coordinates in steps of 5\% of the total mass (see labels; black dotted and dashed lines) and square of the radius of gyration $r_\mathrm{g}^2$ (blue solid line) as a function of time. The black solid line indicates the stellar surface.}
\label{fig:internal-restruct}
\end{figure}

During the thermal relaxation, the merger product first approaches critical surface rotation before the model spins down rapidly (see Fig.~\ref{fig:hrd} in the main text and Fig.~\ref{fig:internal-restruct}). As described in the main text, this spin-down is not driven by angular momentum loss from the star but rather by internal readjustments that change the radius of gyration and thereby the surface rotational velocity. Right after the merger, the core of the merger is hotter and denser than in full equilibrium and the deposited energy from the merger bloats the star leading to the fast thermal expansion of the envelope. At the same time when the envelope contracts, the core expands (Fig.~\ref{fig:internal-restruct}).

The internal magnetic field keeps the star close to solid-body rotation such that the total angular momentum $J$ of the star is $J=r_\mathrm{g}^2 M R_{*}^2 \Omega_{*}$ with $r_\mathrm{g}$ the radius of gyration, $M$ the stellar mass, $R_{*}$ the stellar radius and $\Omega_{*}$ the angular velocity. For constant angular momentum $J$ and mass $M$, the surface rotational velocity evolves according to $v_\mathrm{rot}\, {\propto}\, (r_\mathrm{g}^2 R_{*})^{-1}$ (see Eq.~\ref{eq:vrot}). In the contraction phase when the star spins down (about $10^2\text{--}10^4\,\yr$ after the merger), the radius decreases by a factor of 4 while $r_\mathrm{g}^2$ increases by a factor of 20, fully explaining the observed spin-down of the merger product by a factor of about 5 (Fig.~\ref{fig:internal-restruct}).

\section{Movie of the magnetic-field amplification in the merger}\label{sec:movies}

Supplementary Video S1 illustrates the evolution of the magnetic-field strength from the onset of the merger until the end of the computation.

\vspace{0.5cm}
\noindent {\bf Supplementary Video S1:}
\textbf{Magnetic field in the orbital plane.}
Similar to Fig.~\ref{fig:rho-b-evolution}\textbf{g}--\textbf{i}, in this movie the evolution of the absolute magnetic-field strength is colour-coded while the geometry of the magnetic field is
visualised using line-integral convolution. This illustrates that the magnetic field is initially amplified locally before it organises itself on larger scales.

\end{document}